\documentstyle[twoside,fleqn,espcrc2]{article}

\title{Improvements of the local bosonic algorithm.}

\author{B. Jegerlehner\address{Max-Planck-Institut f\"ur Physik\\
                  F\"ohringer Ring 6, 80807 M\"unchen, Germany}}
       
\def\tint{\tau_{\rm int}}
\def\hq{\hat Q}
\def\tq{\tilde Q}
\def\hc0{\tilde c_0}

\def\umu{U(x,\mu)}

\def\pa{\phi(x)}
\def\pb{\phi(x+\hat\mu)}

\begin{document}

\begin{abstract}
We report on several improvements of the local bosonic algorithm proposed by
M. L\"uscher. We find that preconditioning and over-relaxation works very
well. A detailed comparison between the bosonic and the Kramers-algorithms 
shows comparable performance for the physical situation examined.
\end{abstract}

\maketitle

\section{INTRODUCTION}
Since its proposal in \cite{ml1}, the local bosonic fermion method has been
thoroughly investigated and improved \cite{ml2,bj1,bj2,bj3,f1,f2,f3,wos,mo,f4}. 
We report here on some improvements, especially the efficient implementation
of preconditioning techniques, and give more detailed information on the
comparison of this algorithm to the Kramers algorithm 
presented in \cite{kjhere}. As in \cite{ml2}, all investigations were made with
the gauge group $SU(2)$.

\section{PRECONDITIONING}

\begin{table*}[htb]
\begin{center}
\begin{tabular}{|c|l|l|l|l|l|l|}
\hline
$n,\epsilon$ & $c_M$ & Updating & $\langle \Box \rangle$ & $\langle {\hat\lambda}_{\rm min}\rangle$ & $\langle \hat \lambda_{\rm max} \rangle$
& $\tau_{\rm int}(\Box)$ \\
\hline
0.0017 & 1.1  & HOh,non-pre & 0.4715(12) & 0.00321(8)  & 0.6485(1) & 225(62) \\
50     &      & HOh         & 0.4743(24) & 0.00614(18) & 0.2617(1) &196(52) \\
0.0045 &      & HOh         & 0.4707(15) & 0.00624(16) & &45(10)  \\
30     &      & HOoOo       & 0.4704(11) & 0.00609(8)  & &15(4) \\
0.015  & 0.6  & HOoOo       & 0.4710(16) & 0.00612(11) & 0.8797(3) & 16(5) \\
16     & 0.58 & HOoOo       & 0.4692(13) & 0.00617(6)  & 0.9414(3) & 23(5) \\
\hline
\end{tabular}
\end{center}
\caption{Autocorrelation times in units of $1000 Q\phi$-multiplications
on the $4^3\times 8$ lattice for $\beta=1.75$
and $K=0.165$. $n$ and $\epsilon$ are chosen so that $\delta = 0.03$.
All except the first run are preconditioned. 
The letters in the third column
give the type and order of sweeps used per iteration, where H is a bosonic
heatbath, O a bosonic over-relaxation and h and o the gauge updates. 
\label{tab48pre}}
\end{table*}

Even-odd preconditioning is known to work very well for iterative solvers. It
is therefore straightforward to try to apply this technique to the local
bosonic algorithm. We quickly recall that the hermitean dirac operator 
can be written as
\begin{equation}
Q = c_0 \gamma_5 \left( 
\begin{array}{cc}
1 & -K D_{eo} \\
-K D_{oe} & 1 
\end{array} \right) .
\end{equation}
$c_0 = [c_M(1+8K)]^{-1}$ with $c_M > 1$ is a normalization constant so that $\|Q\| \le 1$.
Using the following identity, 
\begin{equation}
\det \left( 
\begin{array}{cc}
  A & B \\
  C & D 
\end{array} \right) = \det A \det \left( D - C A^{-1} B \right) ,
\label{detform}
\end{equation}
we can immediately write down the preconditioned matrix
\begin{equation}
\hat Q = \tilde c_0 \gamma_5 (1-K^2 D_{oe} D_{eo}) 
\end{equation}
with $\tilde c_0 = [c_M(1+64K^2)]^{-1}$ and $\|\hat Q\| \le 1$. This matrix,
however, contains next-to-nearest neighbour interactions. Since it appears
squared in the local bosonic action, the local updates become rather 
complicated. The local bosonic approximation to the fermion determinant is
\begin{eqnarray}
\det \hq^2 &\approx& [\det P({\hq}^2)]^{-1} 
\nonumber\\
&=& \prod_k [\det (\hq^2 - z_k)]^{-1} \nonumber\\
&=& \prod_k [\det (\hq - r_k) (\hq - r_k^*)]^{-1} ,
\end{eqnarray}
with $P$ a suitably chosen approximation to $1/x$. By applying Equation (\ref{detform}) 
to each single factor in this equation, we get
\begin{eqnarray}\label{bt}
\det\left({\hq} - r_k\right) \propto \hspace{3.8cm} && \nonumber\\
\det \left(
\begin{array}{cc}
\hc0 \gamma_5   & -\hc0 \gamma_5 K D_{eo} \\
- \hc0 \gamma_5 K D_{oe} & \hc0\gamma_5 - r_k
\end{array} \right) .
\end{eqnarray}
Letting $\tq = \hc0/c_0 Q$, we obtain the preconditioned action
\begin{equation}
{\tilde S}_b = \sum_k \phi^\dagger_k (\tq - P_o r_k^*)(\tq - P_o r_k) \phi_k ,
\end{equation}
which is very similar to the original one. $P_o$ denotes the projector on
the odd sites. 

The dynamics of the preconditioned algorithm is investigated in Table \ref{tab48pre}. 
As can be seen, preconditioning saves a factor of about 4 in computer time,
as expected. The analysis of the condition numbers of $Q^2$ and $\hq^2$ show
that the gain is even near $8$. The reason for that is that the largest 
eigenvalue of $\hq^2$ is far below 1. The attempt to exploit this fact by 
adjusting $c_M$ fails because when the largest eigenvalue of $\hq^2$ gets
near 1 the bosonic fields develop slow modes. Detailed investigations show 
that there is a tradeoff between the slow bosonic modes and the number of 
fields. This is however not yet analytically understood.

The idea of applying some reverse transformation to the single factors of the
polynomial can easily be applied to different cases of preconditioning, e.g.
preconditioning with the inverse free fermion matrix, still resulting in a
local algorithm.

\section{HYBRID OVER-RELAXATION}

It is believed that the application of local heatbath and over-relaxation sweeps
in a combination $1:\xi$, where $\xi$ is a suitably defined correlation
length, results in a dynamical critical exponent of $z \approx 1$ 
(see e.g. \cite{wolff}). It was found in \cite{bj1} that the application of
this method to the gauge fields or the bosonic fields only has no effect,
which can be understood as the fact that the coupling of the fields 
dominates the dynamics of the algorithm in the examined case. The application
to both fields however is very efficient, as can be seen from Table 
\ref{tab48hor}. Although this is far from a verification of $z \approx 1$, we
believe that the gain will increase with increasing correlation length.

\begin{table}
\begin{center}
\begin{tabular}{|l|l|l|l|}
\hline
 Updating & $\langle \Box \rangle$ & $\tau_{\rm int}(\Box)$  \\
\hline
 HOh      & 0.4707(15) & 45(10) \\
 HOo      & 0.4701(12) & 30(8)  \\
 HoOo     & 0.4708(14) & 20(4)  \\
 HOoOo    & 0.4704(11) & 15(4)  \\
 HoOoOo   & 0.4702(8)  & 19(5)  \\
\hline
\end{tabular}
\end{center}
\caption{Autocorrelation times in units of $1000 Q\phi$-multiplications.
The run parameters are the same as in the third line of Table 1.
\label{tab48hor}}
\end{table}

\section{COMBINED UPDATE}

In \cite{bj1} an update which updates the bosonic and gauge fields at the 
same time was proposed. The method uses the fact that we can update a gauge 
link $\umu$ according to the effective action after integrating out all
bosonic degrees of freedom $\pa$ and $\pb$ and then refresh $\pa$ and
$\pb$ with heatbath-updates. The hope is, that the effective action induces a
smaller fermionic force on the gauge field, which was found to dominate the
update. Indeed, it is only of order $K^2$ instead of $K$ in the standard
update; however the updating procedure is more complicated what may compensate
for this effect. There also exists no free-field analysis to get a hint about
the dynamical critical exponent of this update. Table \ref{tab48imp} shows that
the update performs comparably to the standard updates. However 
its implementation
is much more difficult and its application to improved actions is probably
not feasible.

\begin{table}
\begin{center}
\begin{tabular}{|c|l|l|l|l|l|l|}
\hline
$n,\epsilon$ & $c_M$ & Upd & $\langle\Box\rangle$ & $\tint(\Box)$ \\
\hline
0.015  & 0.6  & a  & 0.4708(15) & 20(5)  \\
16     &      & b  & 0.4720(14) & 17(4) \\
\hline
\end{tabular}
\end{center}
\caption{\label{tab48imp}Autocorrelation times for the combined update. 
The first line uses one combined over-relaxation step, for the second
a standard bosonic over-relaxation step was added.}
\end{table}

\section{ALGORITHM COMPARISON}

In \cite{kjhere} the comparison between the local bosonic algorithm in its 
hermitean version and the Kramers algorithm has been described. We will
give more detailed information about the runs made and the parameters chosen.
The bosonic parameters were chosen according to the investigations presented
above while the Kramers algorithm was examined in \cite{kj1}.
In Table \ref{tabc1} the parameters for both algorithms are given
as well as the expectation values of the extremal eigenvalues of 
$\hq^2$ (with $c_M$ given as in the Boson section of the table).
In Table \ref{tabc3} the results for the observables are presented.
As can be seen, the parameters were chosen in a way that the observables agree;
it seems however that $\delta$ is slightly too big in the $8^312$ case.

\begin{table}[hbt]
\begin{tabular}{|l|c|c|c|}
\hline 
 & $6^312$ & $8^312$ & $16^4$ \\
\hline
Machine & Q1 & Q1 & QH2 \\
$\langle\lambda_{\rm min}\rangle$ & 0.0115(4) & 0.00539(9) & 0.00478(3) \\
$\langle\lambda_{\rm max}\rangle$ & 0.9386(4) & 0.6100(2) & 0.6961(5) \\
\hline 
Kramers & & & \\
$\epsilon_{md}$ & 0.205 & 0.185 & 0.125 \\
$k$ & 3 & 4 & 5 \\
$\gamma$ & 0.5 & 0.5 & 0.5 \\
\hline
Boson & & & \\
$\epsilon$ & 0.01454 & 0.0061 & 0.0048 \\
$n$ & 18 & 24 & 44 \\
$\delta$ & 2\% & 4\% & 0.38\% \\
$c_M$ & 0.6 & 0.745 & 0.7 \\
\hline
\end{tabular}
\caption{\label{tabc1}Technical parameters and eigenvalues 
of $\hq^2$ for both algorithms.}
\end{table}

\begin{table}[hbt]
\begin{tabular}{|l|l|c|c|c|}
\hline 
& Alg & $6^312$ & $8^312$ & $16^4$ \\
\hline
$\langle\Box\rangle$ & K & 0.5803(2) & 0.5777(3) & 0.5778(1) \\
                     & B & 0.5804(4) & 0.5768(2) & 0.5779(1) \\
$m_\pi$ & K & 1.191(8) & 1.052(8) & 1.003(2) \\
& B & 1.170(12) & 1.044(3) & 1.004(4) \\
$m_\rho$ & K & 1.275(9) & 1.123(11) & 1.060(2) \\
& B & 1.254(14) & 1.112(4) & 1.063(5) \\
$\tau(\Box)$ & K & 480(100) & 979(300) & 540(230) \\
$[$sec$]$ & B & 289(20) & 1781(112) & 990(330) \\
\hline
\end{tabular}
\caption{\label{tabc3}Results for both algorithms. }
\end{table}

\end{document}